\documentclass[preprint2, longabstract]{emulateapj}
\def\hii{\relax \ifmmode {\rm H\,{\sc ii}}\else H\,{\sc ii}\fi}

\shorttitle{Tracing the [FeII]/[NeII] ratio within star forming dwarf galaxies: a \emph{Spitzer} IRS archival survey}

\shortauthors{O'Halloran et al.}

\begin{document}

\title{Tracing the [FeII]/[NeII] ratio and its relationship with other ISM indicators within star forming dwarf galaxies: a \emph{Spitzer} IRS archival study.}

\author{B. O'Halloran\altaffilmark{1},  S. C. Madden\altaffilmark{2},  N. P. Abel\altaffilmark{3}}

\email{boh@physics.gmu.edu}

\altaffiltext{1}{Dept. of Physics \& Astronomy, George Mason University, Fairfax, VA 22030, USA}

\altaffiltext{2}{Service d'Astrophysique, CEA, Saclay, Orme des Merisiers 91191, Gif-sur-Yvette, France}

\altaffiltext{3}{Department of Physics, University of Cincinnati, Cincinnati, OH 45221, USA}

\begin{abstract}

Archival \emph{Spitzer} observations of 41 starburst galaxies that span a wide range in metallicity reveal for the first time a correlation between
the [FeII]/[NeII] 26.0/12.8 $\mu$m ratio and the electron gas density as traced by the 18.7/33.4 $\mu$m [SIII] ratio, with the [FeII]/[NeII] ratio
decreasing with increasing gas density. The correlations of the [FeII]/[NeII] ratio, the PAH peak to continuum strength \& metallicity found in an
earlier paper were confirmed for a larger sample. We also find a strong correlation between the gas density and the PAH peak to continuum strength.
Using shock and photoionization models, we see the driver of the observed [FeII]/[NeII] ratios is metallicity. The majority of [FeII] emission in low
metallicity galaxies may be shock-derived, whilst at high metallicity, the [FeII] emission may be instead dominated by contributions from \hii\ and
in particular from dense PDR regions. However, the observed [FeII]/[NeII] ratios may instead be following a metallicity-abundance relationship, with
iron being less depleted onto grains in low metallicity galaxies - a result that would have profound implications for the use of iron emission lines
as unambiguous tracers of shocks.

\end{abstract}


\section{Introduction}

The presence of massive stars within starbursts undoubtedly plays a huge role in determining the physical conditions within the local interstellar
medium (ISM). High-mass (M$_{init}$ $\geq$ 8M$_{\odot}$) stars formed within typical starbursts drastically affect the dynamics of the surrounding
ISM, through not only the release of ionizing photons which destroy molecular material, but also via supernovae (SNe) which provide thermal and
kinetic energy input into the ISM. The effects of photoionization by high-mass stars on the observed dearth of PAHs have recently been investigated
\citep{mad06,wu06}, while in \cite{oha06} (hereafter Paper I), we examined a sample of 18 galaxies of varying metallicity (from 1/50th to super
solar) with high star formation rates in order to determine whether supernova-driven shocks do indeed play a role in the PAH deficit in low
metallicity environments. If we consider the ratio of the 26 $\mu$m [FeII] line and the 12.8 $\mu$m [NeII] line as a tracer of the strength of
supernova shocks, we found a strong anti-correlation suggesting that strong supernova-driven shocks are indeed present within low metallicity
galaxies. Furthermore, the PAH deficit within these objects may indeed be linked to the presence and strength of these shocks. However, it has not as
yet been conclusively proved that shocks are the dominant process behind the PAH deficit. As already noted, photoionization and perhaps delayed
injection of dust into the ISM \citep{galliano08} each make their own contributions to the PAH deficit, but as yet, it remains unclear as to which of
these three processes is the most dominant. Additionally, is the relationship between the [FeII]/[NeII] ratio and the PAH emission, as indicated in
Paper I, really probing a causal effect? It could very well be that this relationship may be coincidental, or that physical or selection effects may
be the primary drivers for the observed relationship. It would therefore be advisable to further explore the nature of the ISM within such
star-forming environments using additional mid-IR probes in order to expand upon our understanding of the [FeII]/[NeII] ratio, and by extension, its
relationship with key ISM indicators such as PAH strength and metallicity.

\section{Observations and data analysis}

To accomplish this, we have expanded our sample from the 18 objects presented in Paper I by including archival IRS observations of 23 additional
objects, bringing the sample total to 41. The full list of targets is given in Table 1, and listed by increasing metallicity. These galaxies range in
metallicity from extremely low (such as I Zw 18, with $ \it Z/Z_{\odot}$=  1/50) to super-solar metallicity ($\geq 1~\it Z_{\odot}$) galaxies such as
NGC 7714. To differentiate between low and high-metallicity galaxies,  we use a metallicity (12 + log [O/H]) cut-off value of 8.85 - just less than
solar, as per Paper I. The galaxies range widely in morphology from blue compact dwarfs such as I Zw 18 to spirals such as NGC 7714 and IC 342. None
of the galaxies in our sample are known to harbour AGNs. This is important, as PAH destruction can occur close to an AGN due to the hard ionization
environment (e.g. Sturm et al. 2000). In addition [FeII] emission can be elevated in galaxies harbouring AGN. One possible exception is NGC 7714,
which is optically classified  as  a LINER \citep{tho02}. However, the lack of  [NeV] emission at  14 and 24 $\mu$m and  the absence of any evidence
for an obscured AGN by recent Chandra imaging \citep{bran04,smi05} strongly suggests that NGC  7714 is a pure starburst and it is therefore included
in our sample.

We extracted low  and high resolution archival spectral  data from the Short-Low  (SL) (5.2  -  14.5  $\mu$m), Short-High (SH) (9.9 - 19.6 $\mu$m)
and  Long-High (LH) (18.7 -  37.2 $\mu$m) modules  of the {\it Spitzer} Infrared Spectrograph (IRS).   The datasets were derived from a number of
{\it Spitzer}  Legacy, GO and GTO programs released to the  {\it Spitzer} Data  Archive, and  consisted of  either spectral  mapping  or staring
observations. We obtained fluxes for the nuclear positions only from the mapping observations.  All the staring observations were centered on the
galaxy's nucleus. The data were preprocessed by the {\it Spitzer} Science Center (SSC) data reduction  pipeline version  15.3\footnote[4]{{\it
Spitzer} Observers Manual, URL: http://ssc.spitzer.caltech.edu/documents/som/} before being downloaded. Further processing was done within the IRS
BCD-level data reduction package {\it SPICE}, v  2.0.1\footnote[5]{URL: http://ssc.spitzer.caltech.edu/postbcd/spice.html}. Both the low and high
resolution spectra were extracted by {\it SPICE} using the full-aperture extraction method. As the [SIII] emission is extended beyond the extent of
the high resolution slits in particular, we applied the ALCF and SLCF extended source corrections to the spectra as part of the \emph{SPICE}
extraction\footnote[6]{URL: http://ssc.spitzer.caltech.edu/postbcd/doc/spice.pdf}. The slit for the SH and LH modules is too small for background
subtraction to take place and separate SH or LH background observations do not exist for any of the galaxies in this sample. For the SL module,
background subtraction was done using either a designated background file when available or by subtracting the data from one nod position from the
other along the slit. In some cases, the slit was enveloped in the host galaxy and background subtraction could not take place. For both high and low
resolution spectra, the ends  of each order where the noise increases significantly were manually clipped, as were hot pixels.

For the high resolution observations, we required matched extractions in terms of angular extent from both the SH \& LH slits, in order to accurately
derive densities based upon the 18 and 33 $\mu$m [SIII] lines.  The ionization potential of [SIII] is $\sim$35 eV, with the [SIII] emission arising
from gas ionized by young stars for our sample. With a relatively low ionization potential, [SIII] emission may be quite easily extended in spatial
extent beyond the nuclear single-slit pointings for SH and LH. Indeed, comparisons of ISO and IRS data \citep{dudik07} show [SIII] emission within
star-forming AGN that extends beyond the size of both the SH and LH slits. Given that [SIII] emission is extended and we would expect to see a higher
flux in the larger LH slit, any derived [SIII] line ratio will be artificially depressed. To avoid such aperture effects between the SH and LH slits,
we scaled the LH flux by multiplying it by the ratio of the SH/LH slit angular sizes.

As a number of the sample starbursts at high metallicity are known to be dusty systems with large obscuration, the extracted line fluxes were then
corrected for extinction for the sample as a whole. The fluxes and statistical errors, plus the derived line ratios, are presented in Tables 1 and 2
respectively, obtained using the IDL-based analysis package \emph{SMART}, v.6.2.6 \citep{hig04}. In all cases detections were defined when the line
flux was at least 3$\sigma$. For the SL, SH and LH modules the spectral resolution was $\lambda$/$\delta\lambda$ 60-127 and $\sim$600 respectively,
with FWHM in the order of 1.9 x 10$^{-1}$ $\mu$m for the PAH feature in the low-res spectrum, and 2-4 x 10$^{-2}$ $\mu$m for the fine structure lines
in the high resolution spectra. For the absolute photometric flux uncertainty we conservatively adopt 15\%, based on the assessed values given by the
Spitzer Science Center (SSC) over the lifetime of the mission. This error is calculated from multiple observations of various standard stars
throughout the Spitzer mission by the SSC. The dominant component of the total error arises from the uncertainty at mid-IR wavelengths in the stellar
models used in calibration and is systematic rather than Gaussian in nature. The 6.2 $\mu$m PAH strength was determined according to the prescription
adopted in a number of previous studies \citep{rig99,fors04}. While the 7.7 and 8.6 $\mu$m PAH lines are stronger, the 8.6 $\mu$m line can suffer
extinction effects due to a potential silicate absorption feature at 9 $\mu$m, and both of these pose difficulties in determining the continuum
level. For this work, the continuum was determined at the center of the 6.2 $\mu$m feature, using a first order linear fit to the 5.5 and 11.5 $\mu$m
bandpass. The PAH strength was then calculated as the ratio of the 6.2 $\mu$m feature peak intensity to the underlying continuum. We assume a
Gaussian fit while fitting the PAH and fine structure lines with \emph{SMART}.

\section{Results}

In Fig 1, we plot [FeII]/[NeII] vs the 6.2 $\mu$m PAH peak to continuum ratio to check if the relationship between the [FeII]/[NeII] ratio and the
PAH peak to continuum strength first noted in Paper I, using the 6.2 $\mu$m PAH feature instead of the 7.7 $\mu$m, as one can perform continuum
fitting with a higher degree of confidence with the 6.2 $\mu$m feature, still holds for the expanded sample, as noted above. Employing a Spearman
rank correlation analysis \citep{ken76} to assess the statistical significance of this trend yields a correlation coefficient of r$_{s}$ of -0.705
and P$_{s}$ of 2.69 x 10$^{-6}$, confirming the significant anti-correlation between PAH strength and the [FeII]/[NeII] ratio as seen in Paper I. The
Spearman rank correlation technique has the advantage of being non-parametric, robust to outliers and does not presuppose a linear relation. We plot
the [FeII]/[NeII] ratio versus the metallicity in Fig.2, and we again see a strong anti-correlation (r$_{s}$ of -0.777 and P$_{s}$ of 9.36 x
10$^{-7}$) between [FeII]/[NeII] and metallicity as seen in paper I.

\begin{figure}
   \resizebox{\columnwidth}{!}  {\includegraphics{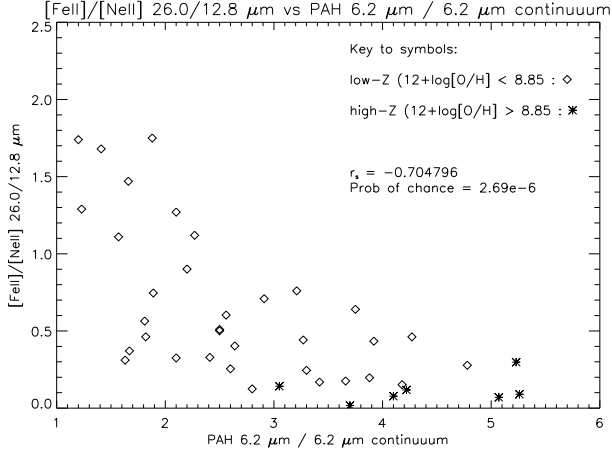}}
\caption{Plot of the [FeII]/[NeII] ratio versus the PAH 6.2 micron peak to continuum ratio for the extended sample. The observed trend confirms
   the relationships between the [FeII]/[NeII] ratio and the PAH strength seen in Paper I.}
\end{figure}

\begin{figure}
   \resizebox{\columnwidth}{!}  {\includegraphics{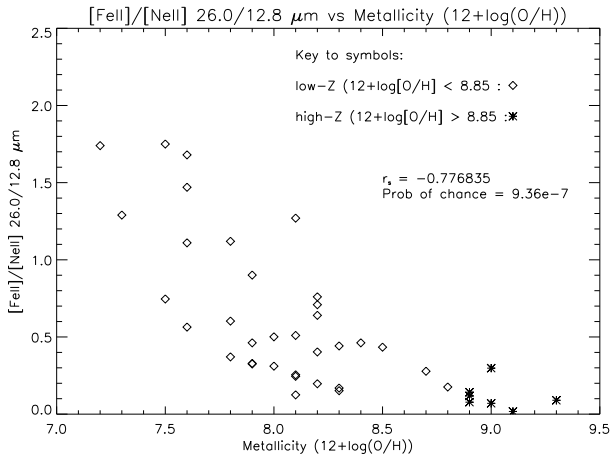}}

   \caption{Plot of the [FeII]/[NeII] ratio versus the metallicity for the extended sample. The observed trend confirms
   the strong relationship between the [FeII]/[NeII] ratio and metallicity seen in Paper I.}
\end{figure}

Based upon what we have already seen from Paper I and from Figs.~1 and 2, if the [FeII]/[NeII] ratio is truly indicative of the strength of
supernova-driven shocks within the extent of the high resolution slits, one would expect the passage of such shocks to affect conditions within the
local ISM in quite a substantive manner. Shocks, in addition to removing dust and PAH from the ISM \citep{reach00,reach06}, should also be adept in
removing gas - one would therefore expect the propagation of intense SNe-driven shocks into the ISM of star forming regions to greatly impact on the
density of the gas. In order to determine how the gas density within these nuclear star forming regions corresponds with the strength of the
supernova-driven shocks, we require a reliable mid-IR tracer to probe the gas density. The [SIII] 18.7/33.4 $\mu$m line ratio provides such a
reliable mid-IR tracer of the gas density, as it is ideal for probing gas (with high critical densities) in the regions surrounding starbursts,
especially as the [SIII] lines are starburst dominated \citep{ver04}. This ratio is sensitive to changes in the density for the 50 $\leq$ n$_{e}$
$\leq$ 10$^{4}$ cm$^{-3}$, but is insensitive to changes in temperature \citep{rigo96}. In Fig. 3, we plot the logarithm of the ratio of the [SIII]
fluxes versus the [FeII]/[NeII] flux ratio. There is a strikingly strong trend between the two line ratios, with high [FeII]/[NeII] values
corresponding to lower [SIII] ratios. Using the Spearman rank correlation analysis, we get (r$_{s}$) of -0.891 between the [SIII] ratio and the
[FeII]/[NeII] with a probability of chance correlation (P$_{s}$) of 7.43 x 10$^{-8}$, indicating a significant anti-correlation. Interestingly, we
see a general decrease in the metallicity of the object with lower [SIII] ratios, corresponding to higher [FeII]/[NeII].

\begin{figure}
   \resizebox{\columnwidth}{!}  {\includegraphics*{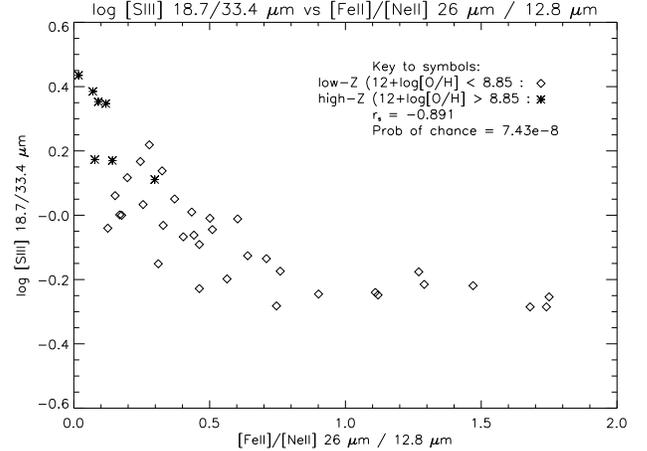}}

   \caption{Plot of the 18.7/33.4 $\mu$m [SIII] ratio as a function of the [FeII]/[NeII] ratio for the extended sample. The [SIII] ratio correlates
   strongly with the [FeII]/[NeII] ratio, with high values of the [FeII]/[NeII] ratio corresponding to low [SIII] ratios and metallicity.}

   \end{figure}

   \begin{figure}
   \resizebox{\columnwidth}{!}  {\includegraphics*{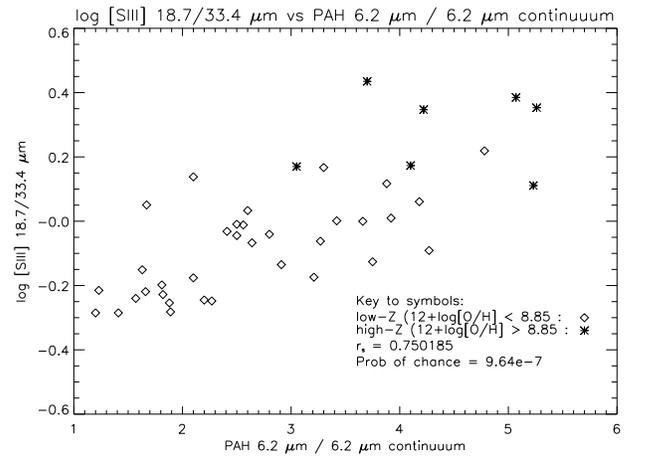}}

   \caption{Plot of the 18.7/33.4 $\mu$m [SIII] ratio as a function of the 6.2 $\mu$m PAH peak to continuum strength for the extended sample. The
   [SIII] ratio correlates quite well with the PAH strength, indicating that the gas density scales with the increased strength of PAH emission.}

   \end{figure}

We again plot the logarithm of the ratio of the [SIII] fluxes in Fig. 4, but this time against the 6.2 micron PAH peak/continuum ratio. As with Fig.
3, we see a strong trend, but this time with the PAH strength increasing with log [SIII] - the PAH strength is increasing with higher gas density.
Again running a Spearman rank correlation analysis, we get r$_{s}$ of 0.750 and P$_{s}$ of 9.64 x 10$^{-7}$, confirming a significant correlation
between log [SIII] and PAH strength. We again see that a general trend exists with PAH strength increasing with metallicity and increasing values of
log [SIII].

\section{Discussion}

\subsection{Can we explain the observed relationships?}

\subsubsection{Are there other factors which may be influencing the observed relationships?}

As we have seen from Figs. 1-4, a series of relationships would seem to exist between the [FeII]/[NeII] ratio, the metallicity, the PAH strength and
the [SIII] ratio, with low values of the [SIII] ratio corresponding with high [FeII]/[NeII] ratios and weak/no PAH emission. However, before we press
ahead with more detailed investigations of the nature of these relationships, it is prudent to first check whether other factors, such as extinction
and aperture effects might be responsible.

\subsubsection{Extinction}

As we have noted previously in Section 2, we corrected the IRS fluxes for extinction, with A$_{V}$ ranging from $\sim$ 2 to 30 mag. To determine the
correction, we used the ratios of $\rm H\beta/H\alpha$ and $\rm Br\beta/Br\alpha$  where available from the literature. Emissivity coefficients from
\cite{storey95} for case B recombination with an electron temperature of 10$^{4}$ K and density of $100~\rm cm^{-3}$ were used to calculate the
intrinsic line ratios for the H recombination lines. We note that while the density used in these calculations is lower than that seen in some of our
sample, we noted little change in the intrinsic line ratios for denser environments. We adopted a \cite{draine89} extinction law above 8.5 $\mu$m,
and the \cite{lutz99} galactic extinction law below this point. As the observed relationships remain post correction, we conclude that extinction can
be ruled out as a driver for the observed correlations. For the optical [SII] lines used lines used as comparisons to the \emph{MAPPINGS} output, we
obtained and used from the literature extinction corrected fluxes where available.

\subsubsection{Aperture effects}

As noted earlier, the ionization potential of [SIII] is $\sim$35 eV, with the [SIII] emission arising from gas ionized by young stars. With a
relatively low ionization potential, [SIII] emission may be quite easily extended in spatial extent beyond the nuclear single-slit pointings for SH
and LH. To compensate, we applied the SLCF and ALCF extended source corrections to the spectra. Additionally, as we are looking to derive a physical
quantity, namely the gas density, over a similarly sized physical area given the extent of the [SIII] emission, it is therefore crucial that we use
[SIII] fluxes derived from slits of equal angular size. To do this, we obtained a fraction of the LH full slit flux by multiplying it by the SH/LH
angular size ratio, as noted in Section 2. The final, corrected fluxes are given in Table 1.

\subsection{Modeling the observed relationships}

Since we can rule out extinction and aperture effects as drivers for the observed correlations, we can be confident that the correlations seen in
Figs. 1-4 are indeed true physical relationships between the strength of the [FeII]/[NeII] ratio and a number of ISM indicators, with the primary
culprit for the observed relationships being the passage of SNe-driven shocks. However, while we have focussed up to now with a supernova-derived
shock origin for the behaviour of the [FeII]/[NeII] ratio, it may not be the only explanation. For example, \cite{izotov06} note sign of strong
depletion of iron onto dust grains, and gradual destruction of those grains on a time scale of a few Myr, based on a survey of metal-poor galaxies
from the 3rd release of the SDSS. Such a process could undoubtedly drastically affect the nature of Fe emission, and by extension the behaviour of
the [FeII]/[NeII] ratio, within our sample - we may instead be probing an abundance-driven relationship and the other observed relationships
presented here would be purely coincidental to the [FeII]/[NeII] ratio. In order to explore if this indeed is the true cause or if shocks alone can
explain the observed relationships, we used shock and standard \hii\ - PDR models in an effort to model the observed [FeII]/[NeII] relationship for a
wide variety of environments. We used both the \emph{MAPPINGS III} photoionization/shock code \citep{sutherland93} and the Cloudy photoionization
code \citep{ferland98} in order to determine the relative proportions to the [FeII] emission from shocks and \hii/PDRs, and by extension, the driving
process behind the behaviour of the observed [FeII]/[NeII] line ratio. In addition to the mid-IR output, we also wished to explore if a similar
relationship is present at optical wavelengths. To this end, we use the ratio of extinction corrected [SII] fluxes at 6716 and 6731~\AA~taken from
the literature and compared the observed ratios with the\emph{ MAPPINGS }output.

\subsubsection{MAPPINGS models}

In an effort to quantify the true nature of the [FeII]/[NeII] ratio, we initially constructed a number of \emph{MAPPINGS} models where density,
metallicity and luminosity were allowed to vary. The theoretical \hii\ region models were generated by the \emph{MAPPINGS III} code, which uses as
input the EUV fields predicted by the stellar population synthesis models \emph{STARBURST99} \citep{leitherer99}.  The photoionization modeling and
shock models carried out with \emph{MAPPINGS III} for this analysis are described in \cite{dopita96} and \cite{kewley01}, and are described briefly
here. To model the \hii\ region spectrum, \emph{MAPPINGS} assumes the metallicity and the shape of the EUV spectrum are defined (through the
\emph{STARBURST99} input), and characterizes the local ionization state by a local ionization parameter:

\begin{equation}
q = \frac{S_{H^{\emph{0}}}}{n}
\end{equation} where \emph{S} is the ionizing photon flux through a unit area, \emph{n} is the local number density of hydrogen atoms and  \emph{q} is the maximum
velocity of an ionization front that can be driven by the local radiation field \citep{dopita00,  kewley02}. We used a 5 Myr continuous
star-formation SED from \emph{STARBURST99} as input to the \emph{MAPPINGS III} code. The parameters of the SED consisted of a Saltpeter IMF with a
power law of 2.35 and a star-formation rate of 1 $\it M_{\odot}$ yr$^{-1}$.

We then constructed a series of plane parallel, isobaric models with \emph{P/k} = 10$^{5}$ cm$^{-3}$ K, where either the gas density, metallicity or
the luminosity for the \emph{STARBURST99} input were allowed to vary, with the other two variable remaining constant. For the \emph{STARBURST99}
input models, the density, metallicity and luminosity were varied as follows:

$\bullet$ \textbf{\emph{Density:}} 10-1000 cm$^{-3}$; metallicity (0.4\emph{Z$_{\odot}$}) and luminosity (10$^{43}$ ergs) remain constant;

$\bullet$  \textbf{\emph{Metallicity:}} 0.05 - 2\emph{Z$_{\odot}$}; density (100 cm$^{-3}$) and luminosity (10$^{43}$ ergs) remain constant;

$\bullet$ \textbf{\emph{Luminosity:}} 10$^{41}$ - 10$^{45}$ ergs; density (100 cm$^{-3}$) and metallicity (0.4\emph{Z$_{\odot}$}) remain constant.

Elemental abundances for 0.5-2.0\emph{Z$_{\odot}$} were adopted for the variable metallicity models, while for the variable density and luminosity
models, 0.4\emph{Z$_{\odot}$} metallicity abundances were used.

Dust physics is treated explicitly through absorption, grain charging and photoelectric heating and the use of a standard MRN  grain size
distribution for both carbonaceous and silicaceous grain types \citep{mat77}. The \emph{MAPPINGS} code took the undepleted solar abundances to be
those of \cite{anders89}. For non-solar metallicities, \emph{MAPPINGS} assumes that both the dust model and the depletion factors are unchanged. All
elements except nitrogen and helium are taken to be primary nucleosynthesis elements. As for the shock models, we use the pre-run shock models from
the \emph{MAPPINGS} website, which are based upon a grid of \emph{n} = 1.0 models with \emph{v} = 100-1000 km/s in steps of 25 km/s. The magnetic
parameter \emph{B = p/$\sqrt{n}$} is fixed to 3 in all calculations. This value corresponds to equipartition between thermal and magnetic pressures
\citep{dopita96}.

 \begin{figure}
   \resizebox{\columnwidth}{!}{\includegraphics*{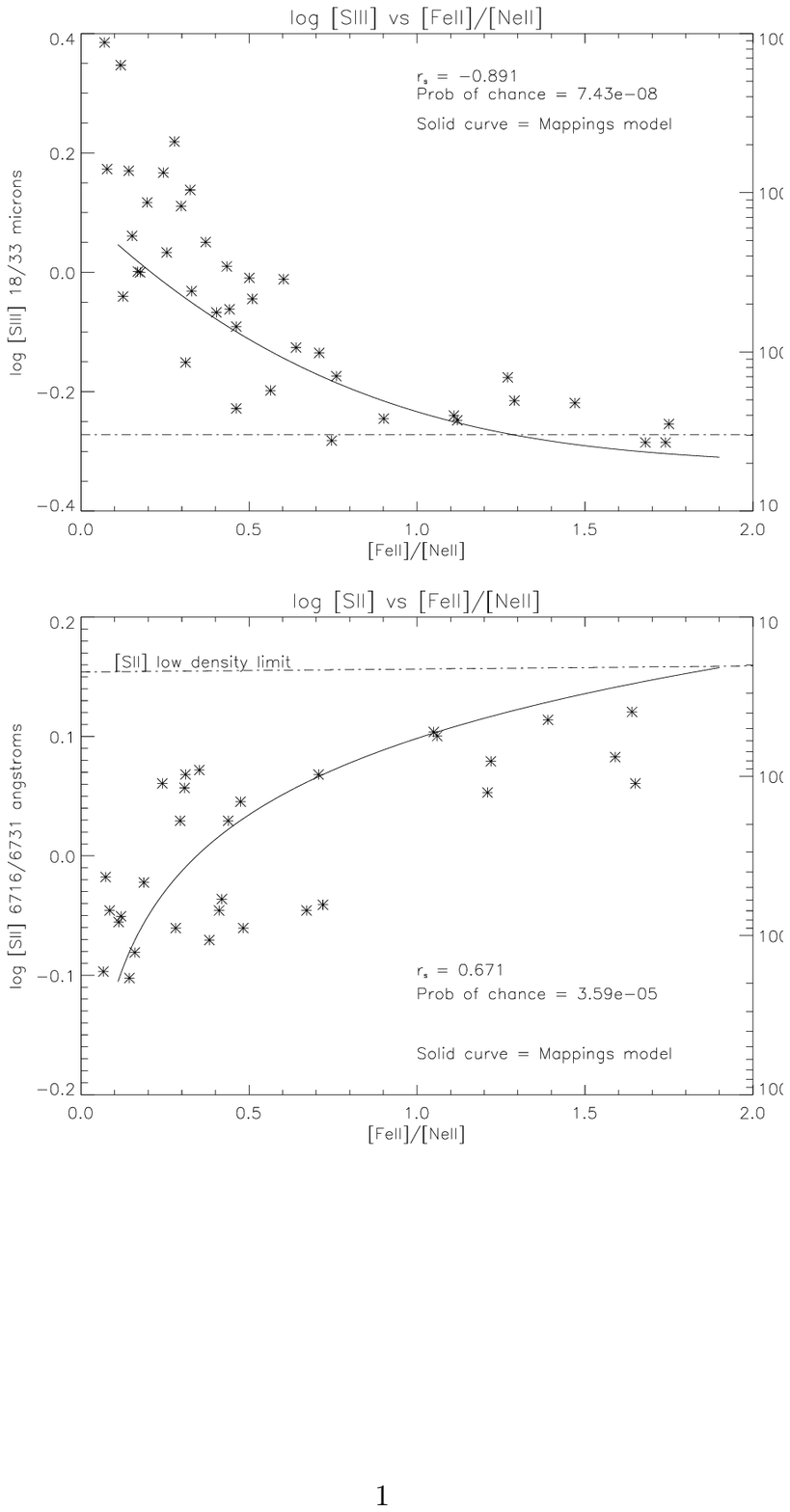}}

   \caption{Plots of the [SIII] (top) and [SII] (bottom) ratios vs [FeII]/[NeII] for the extended galaxy sample, overlaid with model data from \emph{MAPPINGS}. The dot-dash line
   indicates the low density limit.}

  \end{figure}

From the \emph{MAPPINGS} output from the variable metallicity model, we saw that whilst the [SIII] and [SII] values remain relatively constant (and
by extension, the density), the [FeII]/[NeII] ratio varies quite considerably. With the variable density model, interestingly, we saw that
[FeII]/[NeII] does not drop with increasing density contrary to what we observe (Fig. 3), but rather slightly increases, contrary to what we observe.
Finally, for the variable luminosity model, both the [FeII]/[NeII] and unsurprisingly the [SIII] and [SII] values, remained constant. From the
\emph{MAPPINGS} model data, it would seem that the change in metallicity is primarily responsible for the observed drop in the [FeII]/[NeII] ratio.
In order to directly compare the photoionization and shock model output with the IRS and optical data, we then set about defining models that would
replicate the sort of density, metallicity and luminosity parameters typically seen in the sample. Since the low metallicity objects tend to have
lower densities and luminosities, we generated  \emph{MAPPINGS} models that gradually increased the metallicity, density and luminosity, over the
ranges outlined above, moving from low to high metallicity (0.05, 0.2, 0.4 and solar metallicity), low to high density (20-1000 cm$^{-3})$ and low to
high luminosity (10$^{42}$ to 10$^{44}$ ergs). As we increased the metallicity, we concurrently moved to high densities and higher luminosities. As
with the earlier \emph{MAPPINGS} models, we used a 5 Myr continuous star-formation SED from \emph{STARBURST99} as input to the \emph{MAPPINGS III}
code \citep{leitherer99}. The parameters of the SED consisted of a Saltpeter IMF with a power law of 2.35 and a star-formation rate of 1 $\it
M_{\odot}$ yr$^{-1}$. Elemental abundances for 1/20\emph{Z$_{\odot}$}, 2/5\emph{Z$_{\odot}$} and 1\emph{Z$_{\odot}$} were again adopted.

Using the observed [SII] 6716 \& 6731~\AA, [SIII] 18 \& 33 $\mu$m plus the [FeII] 26 $\mu$m and [NeII] 12.8 $\mu$m fluxes, we obtained the respective
ratios and performed best fit line calculations to the [SIII] vs [FeII]/[NeII] and [SII] vs [FeII]/[NeII] model data, where the line fluxes are the
product of both the photoionization and shock model outputs. Using the best fit models, we plotted the model ratio data in conjunction with the
observed IRS and optical data ratios (Fig. 5), with the model data derived from the second $\it MAPPINGS$ run describing very well the behaviour of
both sets of observed line ratios.

\subsubsection{Cloudy models}

The slight increase in the [FeII] emission with density seen from the \emph{MAPPINGS} output poses the following question - where does this increase
originate from? Is this increase due to an increased contribution to the [FeII] emission by dense \hii\ regions and PDRs? If so, what is the level of
this contribution? To resolve this, we used the spectral synthesis code Cloudy to try and determine how much of a contribution \hii\ regions and PDRs
make to the overall [FeII] emission, how this contribution is affected by the gas metallicity (as suggested by \emph{MAPPINGS}), and the effect of
the \hii/PDR contribution has on the overall [FeII]/[NeII] ratio. We used the developmental version of the spectral synthesis code Cloudy, last
described by \cite{ferland98}.  \cite{abel05} and \cite{shaw05} describe recent advances in its treatment of PDR chemistry, while \cite{vanhoof04}
describes the grain physics. As with the \emph{MAPPINGS} models, we used a 5 Myr continuous star-formation SED from \emph{STARBURST99} as input to
Cloudy. The parameters of the SED consisted of a Saltpeter IMF with a power law of 2.35 and a star-formation rate of 1 $\it M_{\odot}$ yr$^{-1}$.

Our model geometry consists of a plane-parallel slab illuminated on one side by a source of UV radiation.  Our calculations start at the hot,
illuminated face of the slab, where all the hydrogen is ionized.  We end our calculations at two locations, when the hydrogen ionization fraction
falls below 1\% (for the pure \hii\ region calculations, see below) and when the fraction of hydrogen in the form of \emph{H$_{2}$}(\emph{(2 $\times$
n(H$_{2}$))/{n(H)$_{total}$})} exceeds 90\%.  Beyond this depth, we expect there to be little contribution to either [FeII] or [NeII] emission.

We parameterize the ionizing continuum with the `ionization parameter', which is the dimensionless ratio of hydrogen ionizing flux to density, \[U =
\frac{Q_H}{4 \pi R^2 N_H c}\] where \emph{Q$_{H}$} is the total number of hydrogen-ionizing photons emitted by the central object per second. For all
calculations, we chose Log[\emph{U}] = -2.5, a value typical in \hii\ regions \citep{veilleux87}. The ionization parameter in \emph{MAPPINGS} is
related to the one used in Cloudy by the equation:

\begin{equation}
U = \frac{q}{c}
\end{equation}

We also include cosmic rays in our calculations.  Primary and secondary cosmic ray ionization processes are treated as described in Appendix C of
\cite{abel05}, with an assumed cosmic ray ionization rate of 5 $\times$ $10^{-17}$ s$^{-1}$.  Cosmic rays are a significant heating source and also
drive the ion-molecule chemistry deep in the molecular cloud.  However, since our calculations stop short of the molecular cloud, for our models
cosmic rays are only of secondary importance. Our calculations include the major ionization processes that can affect the ionization structure. This
includes all stages of ionization for the lightest 30 elements.

Dust is known to play an important role in both \hii\ regions and PDRs \citep{draine03}.  We self-consistently determine the grain temperature and
charge as a function of grain size and material, for the local physical conditions and radiation field.  This determines the grain photoelectric
heating of the gas, an important gas heating process, as well as collisional energy exchange between the gas and dust.  We also treat stochastic
heating of grains as outlined in \cite{guhathakurta89}, which can affect the dust continuum shape.  We include grain charge transfer as a general
ionization - recombination process, as described in Appendix B of \cite{abel05}.  The rates at which H$_{2}$ forms on grain surfaces is derived using
the temperature and material-dependent rates given in \cite{cazaux02}.  The assumed grain size distribution is representative of the star-forming
regions.  The ratio of total to selective extinction, \emph{R$_{V}$} = \emph{A$_{V}$}/(\emph{A$_{B}$}-\emph{A$_{V}$}), is a good indicator of the
size distribution of grains \citep{cardelli89}. \cite{calzetti00} derived an average value for \emph{R$_{V}$} $\sim$ 4.3.  We therefore use the
\emph{R$_{V}$} = 4 grain size distribution given in \cite{weingartner01}.  We also include size-resolved PAHs in our calculations, with the same size
distribution used by \cite{bakes94}.  The abundance of carbon atoms in PAHs that we use, \emph{n$_{C}$}(PAH)/\emph{n$_{H}$}, is 3 $\times$ 10$^{-6}$.
PAHs are thought to be destroyed by hydrogen ionizing radiation and coagulate in molecular environments (see, for instance, \cite{omont86}).  We
model this effect by scaling the PAH abundance by the ratio of H$^{0}$/H$_{tot}$(n$_{C}$(PAH)/n$_{H}$ = 3 $\times$ 10$^{-6}$
[\emph{n}(H$^{0}$)/\emph{n}(H$_{tot}$)]).

The assumed gas-phase and dust abundances are extremely important in our calculations. We also assume gas-phase abundances based on an average of the
abundances in the Orion Nebula derived by \cite{baldwin91}, \cite{rubin91}, and \cite{osterbrock92}.  The abundances by number are He/H = 0.095, C/H=
3 $\times$ 10$^{-4}$; O/H= 4 $\times$ 10$^{-4}$, N/H= 7 $\times$ 10$^{-5}$, and Ar/H= 3 $\times$ 10$^{-6}$.  We also assume default grain abundance
of A$_{V}$/\emph{N}(H$_{tot}$) = 5 x 10$^{-22}$ mag cm$^{2}$.  We then scale the metal and dust abundance such that the final value of \emph{Z} =
12+log[\emph{O/H}] ranges from 7 to 10, in increments of 1 dex.

   \begin{figure}
   \resizebox{\columnwidth}{!}  {\includegraphics*{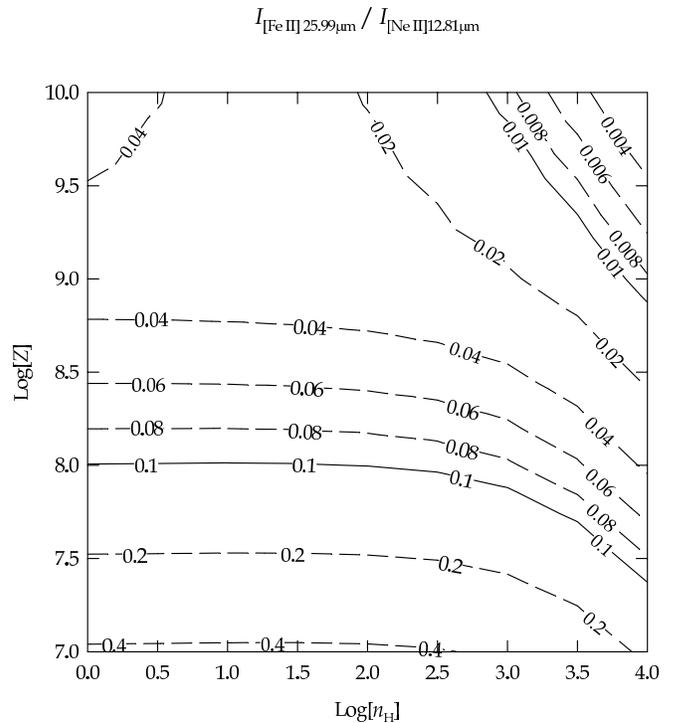}}

   \caption{ Cloudy photoionization model output of the [FeII]/[NeII] ratio as a function of density and \emph{Z}, with contributions to the[FeII] emission by \hii\ regions only.}

 \end{figure}

  \begin{figure}
   \resizebox{\columnwidth}{!}  {\includegraphics[angle=0]{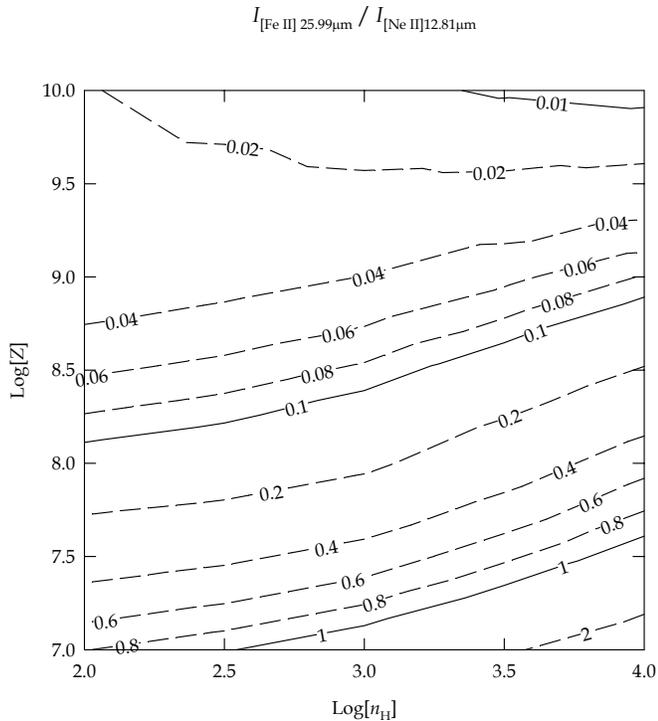}}

   \caption{ Cloudy photoionization model output of the [FeII]/[NeII] ratio as a function of density and \emph{Z}
   with contributions to the [FeII] emission from both \hii\ and PDR regions.}

 \end{figure}

There is one major exception to our assumed abundances, and that is the \emph{Fe/H} ratio.  \cite{rodriguez05} discuss the variation in \emph{Fe/H}
with \emph{Z}, and as noted earlier, \cite{izotov06} also determined the variation in elemental abundances with \emph{Z} from the Sloan Digital Sky
Survey, including Ne and Fe. For these two elements, \cite{izotov06} provide the following equations to derive their respective abundances (for this
work, we use the notation \emph{Z} instead of \emph{X}, as used in \cite{izotov06}:

\begin{equation}
Log(Ne/O) = 0.097\emph{Z}- 1.542
\end{equation}

and

\begin{equation}
Log(Fe/O) = -0.606\emph{Z} + 2.994
\end{equation}

where \emph{Z} = 12 + Log(\emph{O/H}), we find that:

\begin{equation}
Log \left[\frac{Ne}{Fe}\right]   = -4.536 + 0.703Z.
\end{equation}

In all calculations, we scale the Fe abundance so that equation 3 is always obeyed.

The relationship between density, temperature, and pressure is determined by the assumed equation of state.  We perform two sets of calculations, one
where we stop the calculation at the hydrogen ionization front and one where we include the PDR.  For the pure \hii\ region models we assume constant
density, with the density ranging from Log[\emph{n}$_{H}$] 0.5 to Log[\emph{n}$_{H}$] = 4, in increments of 0.5 dex.  For the \hii\ + PDR
calculations, we assume constant pressure, with thermal and radiation pressure being the dominant contributors to the total pressure.  We do not
include turbulent pressure in the equation of state, although we assume a small turbulent broadening of 1 km s$^{-1}$ for its effects on line optical
depths.  Such an equation of state neglects shocks or other time-dependent effects.  In the constant pressure scenario, we specify \emph{n}$_{H}$ at
the illuminated face, then solve for density based how the pressure contributions change with depth.  Our densities range over Log[\emph{n}$_{H}$] =
2 to 4, in increments of 0.5 dex.  Overall, our combination of \emph{n}$_{H}$ and \emph{Z} represent 52 calculations.

In Fig. 6, we plot the [FeII]/[NeII] line ratio as a function of $\it Z$ and density for the \hii\ region only. In Fig. 7, we again plot the
[FeII]/[NeII] line ratio as a function of $\it Z$ and density, but this time including the PDR in addition to the \hii\ region. As one can see, the
relative \hii\ contribution dominates over the PDR contribution at low densities (and metallicities), with the PDR contribution becoming more
dominant as one moves to higher densities ($\sim$ 10$^{3}$ cm$^{-3}$) and metallicities. Previous IR studies of starburst \hii/PDR regions
\citep{carral94,lord96} support this scenario. For very dense \hii\ regions where the PDRs are irradiated by intense FUV, and thus PDR-derived [FeII]
emission is dominant.  Correspondingly at lower densities ($\sim$10$^{2}$ cm$^{-3}$), the \hii\ regions are larger, the PDRs lie further from the
stars, and the resultant lowered FUV fluxes and densities do not excite [FeII] in the PDR, and thus the \hii\ region dominates the \hii/PDR
contribution to [FeII] production for such environments.

\subsubsection{What is driving the Fe/Ne relationship?}

From the $\it MAPPINGS$ and $\it Cloudy$ simulations and the comparisons with the IRS and optical data (Fig. 5), it would seem that metallicity is
the primary driver for the behaviour of the [FeII]/[NeII] ratio, as metallicity determines the origin (shocks vs \hii\ and/or PDRs) and strength of
the [FeII] emission. We can reproduce the observed [FeII]/[NeII] dependence on \emph{Z} if we assume the Izotov variation of \emph{Fe/Ne} with
\emph{Z}. Cloudy reproduces the [FeII]/[NeII] ratio decrease with increases in density (for the pure H II region only) because of the following: the
critical density for electron excitation of [Fe II] is 1.2 x 10$^{4}$ cm$^{-3}$, while for [Ne II] it is 5 x 10$^{5}$ cm$^{-3}$. The ratio of [Fe
II]/[NeII] thus depends on the ratio:

\[\frac{(1 + n_{critical}(neon)/n(\emph{H}))} {(1 + n_{critical}(iron)/n(\emph{H}))}\]

In the low density limit, this ratio is $\sim$42.  However, for a density of 1 x 10$^{4}$ cm$^{-3}$, which is the high end of our parameter space,
this ratio is only $\sim$20 because [Fe II] emission is suppressed by the higher density. The reason the ratio increases with density in Fig. 7 (\hii
+ PDR) is because [Fe II] is excited by atomic hydrogen in the PDR in addition to electrons in the \hii\ region.  The critical density of [Fe II]
with H atoms is $\sim$1 x 10$^{6}$ cm$^{-3}$. Therefore [Fe II] emits readily in the PDR, while [Ne II] does not emit because all the neon is
neutral. However, we do note that the \hii\ and PDR contribution to the [FeII] emission is surpassed by the contribution from shocks at the lowest
metallicities and densities, based on comparison with the \emph{MAPPINGS} output and the observed data - perhaps only $\sim$10\% of the [FeII]
emission comes from \hii\ and PDRs with that contribution rising to $\sim$80-90\% as one moves to higher metallicity. As we move to roughly solar
metallicity, it would seem likely that the dominant mechanism for production of [FeII] (and thus the behaviour of the [FeII]/[NeII] ratio at high
metallicities) is from \hii\ and PDRs in combination rather than from SNe shocks.

\subsubsection{Can we tie in the behaviour of the gas density with the [FeII]/[NeII] ratio, metallicity and PAH strength?}

While abundances may play a role in the overall behavior of the [FeII]/[NeII] ratio, the shock scenario would seem to fit the observed variation of
the gas density and the PAH ratio with metallicity. Assuming that shocks are indeed behind the observed correlations, the following scenarios would
seem to account for the relationship between the [FeII]/[NeII] and [SIII] ratios and the PAH strength. In very low metallicity dwarf starburst
environments, such as I Zw 18, the observed gas density is phenomenally low, in the order of $\sim$20-100 cm$^{-3}$. From Fig. 3 and the models, it
would seem that low density correlates strongly with the presence of strong supernovae shocks, a harsh radiation environment which is lacking in
populations of dust and PAH, as noted by the lack of PAH emission (Figs. 1 and 4) and generally flat slopes of the dust continua. It would seem
likely therefore that we are probing the tenuous gas within a superbubble cavity that has been carved by the cumulative effects of numerous
supernova-driven shocks. Previous surveys of low metallicity dwarf galaxies have found evidence for such superbubbles
\citep{strickland00,calzetti04,ott05}, with ionized gas expanding at a rate of several tens of km/s driven by mechanical energy from starbursts
deposited into the ISM \citep{calzetti04}.

For high metallicity environments, the picture becomes somewhat more complicated. Such objects are of course still the sites of high levels of
massive star formation - the high metallicity starburst M82 is after all the template for starburst activity - but we see a much more benign
environment for the survivability of dust and PAH. A number of new factors play a role in affecting star formation and its effects on the local
environment. First of all, from the Cloudy and \emph{MAPPINGS} models plus the observed data, we note that the dominant contribution to the [FeII]
flux no longer comes from shocks, but rather from \hii\ regions and from PDRs in particular - the lack of shocks propagating into the ISM provides a
more benign, denser environment for the build-up of dust and PAH. Superbubbles are present as in low metallicity galaxies, however supernova-driven
shocks are no longer the dominant mechanism driving the expansion but rather stellar winds from massive stars. As a consequence, less mechanical
energy is available to drive breakout and outflow and the superbubble can remain enclosed, especially where the gravitational potential of the galaxy
is steep \citep{skill95,burkert04,calzetti04}. Given the inability of weaker shocks to drive out material and clear the local ISM, it is not
surprising that we should see a buildup of metals, and the survivability of dust and PAH formed locally.

\subsubsection{Are abundances driving the Fe/Ne relationship?}

Whilst the SNe-driven shock hypothesis is convincing in terms of the enhanced [FeII] emission and corresponding low PAH emission, it is the not the
only plausible explanation for the observed correlation. As we have already noted, from \cite{izotov06} it has been suggested that the observed
levels of [FeII] emission may instead be abundance driven. Using equations 1 \& 2 to derive how the \emph{Fe/Ne} ratio varies with metallicity, we
see that the \emph{Fe/Ne} ratio decreases with increasing \emph{Z} in a fashion very similar to the [FeII]/[NeII] vs.~\emph{Z} correlation,
suggesting that the variation in the [FeII]/[NeII] emission ratio with \emph{Z} may be a consequence of varying \emph{Fe/Ne} abundance with \emph{Z}.
Indeed, the slope of \emph{Fe/Ne} vs. \emph{Z} from \cite{izotov06} is very similar to the slope of the best fitting line from Fig 2 (m = -0.68) -
our observational dataset points to this being a possibility, and the \emph{MAPPINGS} and Cloudy models further lend support to this scenario.

However given the similarity in terms of behaviour of our observational dataset with the \cite{izotov06} results, it begs the question - what may be
causing the change in \emph{Fe/Ne} with \emph{Z}? Since neon is not depleted \citep{juett06,yao06}, the observed relationship would then be due to
variations in \emph{Fe}/\emph{H} rather than \emph{Ne}/\emph{H}. So what is causing the iron depletion? Since iron is known to be depleted onto
grains \citep{juett06,yao06}, the increased variation in \emph{Fe}/\emph{H} would suggest that there is either more grain destruction, or less grain
formation, as one moves to lower metallicities.

If this is correct, the reason for the observed [FeII]/[NeII] with \emph{Z} trend is that Fe is less depleted in low metallicity galaxies. This would
have a profound impact on ISM studies, as the usefulness of iron emission as a pure indicator of shocks may be compromised. However, this is not
conclusive, and we need more observations to deduce for certain the physical process which is controlling the [FeII] emission. In order to
definitively figure out if shocks are truly responsible for the [FeII] emission at low metallicity, we require high resolution IR spectroscopy to
unambiguously determine the presence of shocks within the velocity profiles. The upcoming availability of such high resolution spectroscopic data
from {\it Herschel} and {\it SOFIA} will be crucial to determine the true nature of iron emission at low metallicity and we look forward to
furthering this line of investigation in a future paper.

\section{Conclusions}

Using archival \emph{Spitzer} observations of 41 starburst galaxies that span a wide range in metallicity, we found a correlation between the ratio
of emission line fluxes of [FeII] at 26 $\mu$m and [NeII] at 12.8 $\mu$m and the electron gas density as traced by the 18.7/33.4 $\mu$m [SIII] ratio,
with the [FeII]/[NeII] flux ratio decreasing with increasing gas density. We also find a strong correlation between the gas density and the PAH peak
to continuum strength. The correlation of the [FeII]/[NeII] ratio and the PAH peak to continuum strength found in \cite{oha06} was confirmed for a
larger sample. Using shock and photoionization models, we see that metallicity is the primary driver for the observed behaviour of the [FeII]/[NeII]
ratio. It may very well be that the majority of [FeII] emission at low metallicity may be shock-derived, whilst at high metallicity, the [FeII]
emission is dominated by contributions from \hii\ and in particular, from PDR regions, and that at higher metallicity shocks may not play as
significant a role in removing gas, PAH and dust from the ISM, unlike in low metallicity systems. However, the observed [FeII]/[NeII] emission may
instead be following a metallicity-abundance relationship, with the iron being less depleted in low metallicity galaxies, a result that would have
profound implications for the use of Fe emission lines as unambiguous tracers of shocks. Follow up high resolution spectroscopic observations will be
required to determine if this is indeed the case, or if the detected iron emission is truly tracing emission from the passage of shocks.

\acknowledgments

We are grateful to S. Satyapal for her comments and guidance. BO'H gratefully acknowledges financial support from NASA grant NAG5-11432. NPA
acknowledges National Science Foundation support under Grant No. 0094050 , 0607497 to The University of Cincinnati. This work is based on archival
data obtained with the \emph{Spitzer} Space Telescope, which is operated by the Jet Propulsion Laboratory, California Institute of Technology under a
contract with NASA. The IRS was a collaborative venture between Cornell University and Ball Aerospace Corporation funded by NASA through the Jet
Propulsion Laboratory and Ames Research Center. \emph{SMART} was developed at Cornell University and is available through the \emph{Spitzer} Science
Center at Caltech. The \emph{SPICE} reduction package has been developed by and released through the \emph{Spitzer} Science Center.

\begin{deluxetable*}{lcrrrrr}
\tabletypesize{\scriptsize} \tablecaption{IRS flux data.} \tablewidth{0pt} \tablehead{\colhead{Target}    &
 \colhead{Metallicity}    &
 \colhead{F([NeII])}     &
 \colhead{F([SIII]) }    &
 \colhead{F([FeII]) }    &
 \colhead{F([SIII]) }    \\
 \colhead{}   &
 \colhead{12+log(O/H)}   &
 \colhead{@ 12.8 $\mu$m}   &
 \colhead{@ 18.7 $\mu$m}   &
 \colhead{@ 26.0 $\mu$m}   &
 \colhead{@ 33.4 $\mu$m}   \\
 \colhead{}   &
 \colhead{}   &
 \colhead{\it (x 10$^{-20}$ W cm$^{-2}$)}   &
 \colhead{\it (x 10$^{-20}$ W cm$^{-2}$)}   &
 \colhead{\it (x 10$^{-20}$ W cm$^{-2}$)}   &
 \colhead{\it (x 10$^{-20}$ W cm$^{-2}$)}   \\
 \colhead{(1)}   &
 \colhead{(2)}   &
 \colhead{(3)}   &
 \colhead{(4)}   &
 \colhead{(5)}   &
 \colhead{(6)}    }

  \startdata
I Zw 18   & 7.2 & 0.030$\pm$0.003 & 0.006$\pm$0.001 & 0.051$\pm$0.015 & 0.011$\pm$0.006\\
SBS 0335-052 & 7.3 & 0.014 $\pm$ 0.003 & 0.013$\pm$0.003 & 0.018$\pm$0.006 & 0.021$\pm$0.008 \\
HS 2236+1344 & 7.5 & 0.032 $\pm$ 0.002 & 0.063$\pm$0.008 & 0.056$\pm$0.016 & 0.112$\pm$0.011 \\
SDSS J0519+0007 & 7.5 &  0.207$\pm$0.014  & 0.048$\pm$0.014 & 0.155$\pm$0.056 & \\
SBS 1415+437 & 7.6 & 0.068$\pm$0.013  & 0.026$\pm$0.007 & 0.076$\pm$0.024 & 0.045$\pm$0.021 \\
HS 0837+4717 & 7.6 &  0.121$\pm$0.010  & 0.020$\pm$0.005 & 0.177$\pm$0.036 & 0.033$\pm$0.011 \\
Tol 65 & 7.6 & 0.024$\pm$0.013  & 0.011$\pm$0.002 & 0.040$\pm$0.005 & 0.022$\pm$0.008 \\
HS 1442+4250 & 7.6 & 0.226$\pm$0.010  & 0.024$\pm$0.004 & 0.127$\pm$0.041 & 0.038$\pm$0.004 \\
Mrk 25 & 7.8 & 1.074$\pm$0.156  & 0.847$\pm$0.104 & 0.647$\pm$0.128 & 0.870$\pm$0.026 \\
SBS 1030+583 & 7.8 & 0.075$\pm$0.014  & 0.167$\pm$0.011 & 0.028$\pm$0.007 & 0.149$\pm$0.043 \\
Mrk 209 & 7.8 & 0.068$\pm$0.017  & 0.070$\pm$0.024 & 0.077$\pm$0.020 & 0.124$\pm$0.024 \\
Mrk 36 & 7.9 & 0.281$\pm$0.071  & 0.051$\pm$0.010 & 0.130$\pm$0.024 & 0.085$\pm$0.006 \\
SBS 0917+527 & 7.9 & 0.228$\pm$0.058  & 0.115$\pm$0.024 & 0.074$\pm$0.007 & 0.084$\pm$0.008 \\
SBS 1152+579 & 7.9 & 0.928$\pm$0.300  & 0.071$\pm$0.037 & 0.305$\pm$0.018 & 0.076$\pm$0.024 \\
Mrk 170 & 7.9 & 0.053 $\pm$ 0.015  & 0.033$\pm$0.002 & 0.048$\pm$0.006 & 0.058$\pm$0.009 \\
UM 448 & 8.0 & 1.430$\pm$0.018  & 0.397$\pm$0.031 & 0.442$\pm$0.084 & 0.563$\pm$0.099 \\
SBS 0946+558 & 8.0 & 0.053$\pm$0.015  & 0.034$\pm$0.022 & 0.026$\pm$0.003 & 0.034$\pm$0.006 \\
II Zw 40 & 8.1 & 0.452$\pm$0.061  & 1.375$\pm$0.026 & 0.115$\pm$0.031 & 1.274$\pm$0.304 \\
Mrk 930 & 8.1 & 0.219$\pm$0.074  & 0.170$\pm$0.019 & 0.112$\pm$0.026 & 0.189$\pm$0.056 \\
Mrk 996 & 8.1 & 0.297$\pm$0.022  & 0.271$\pm$0.044 & 0.073$\pm$0.011 & 0.184$\pm$0.046 \\
Tol 1924-416 & 8.1 & 0.464$\pm$0.039  & 0.247$\pm$0.050 & 0.058$\pm$0.001 & 0.271$\pm$0.073 \\
II Zw 70 & 8.1 & 0.143$\pm$0.012  & 0.178$\pm$0.041 & 0.182$\pm$0.053 & 0.268$\pm$0.075 \\
NGC 4670 & 8.2 & 0.482$\pm$0.173  & 0.240$\pm$0.039 & 0.309$\pm$0.042 & 0.321$\pm$0.029 \\
Mrk 450 & 8.2 & 0.125$\pm$0.009  & 0.083$\pm$0.029 & 0.050$\pm$0.015 & 0.096$\pm$0.059 \\
Mrk 5 & 8.2 & 0.106$\pm$0.013  & 0.076$\pm$0.020 & 0.075$\pm$0.019 & 0.103$\pm$0.027 \\
Mrk 1329 & 8.2 & 0.305$\pm$0.015  & 0.257$\pm$0.054 & 0.060$\pm$0.002 & 0.196$\pm$0.036 \\
NGC 5253 & 8.2 & 0.572$\pm$0.075  & 1.172$\pm$0.039 & 0.434$\pm$0.100 & 1.749$\pm$0.249 \\
IC 342 & 8.3 & 34.239$\pm$5.285  & 9.902$\pm$0.156 & 5.793$\pm$0.025 & 9.872$\pm$1.771 \\
Haro 3 & 8.3 & 2.016$\pm$0.084  & 1.431$\pm$0.064 & 0.306$\pm$0.042 &  1.243$\pm$0.090 \\
UM 311 & 8.3 & 0.126$\pm$0.010  & 0.0873$\pm$0.021 & 0.056$\pm$0.009 & 0.101$\pm$0.042 \\
Mrk 1499 & 8.4 & 0.150$\pm$0.014  & 0.319$\pm$0.008 & 0.069$\pm$0.017 & 0.393$\pm$0.018 \\
UGC 4274 & 8.5 & 0.710$\pm$0.030  & 0.402$\pm$0.036 & 0.308$\pm$0.079 & 0.393$\pm$0.093 \\
NGC 7793 & 8.7 & 0.318$\pm$0.077  & 0.295$\pm$0.010 & 0.088$\pm$0.025 & 0.178$\pm$0.069 \\
NGC 4194 & 8.8 & 8.464$\pm$0.169  & 2.151$\pm$0.042 & 1.491$\pm$0.336 & 2.151$\pm$0.466 \\
NGC 253 & 8.9 & 199.753$\pm$1.235  & 42.928$\pm$0.780 & 28.433$\pm$1.154 & 29.012$\pm$0.416 \\
He 2-10 & 8.9 & 24.671$\pm$1.677  & 9.034$\pm$0.156 & 2.300$\pm$0.091 & 6.072$\pm$0.990 \\
NGC 7714 & 8.9 & 8.546$\pm$0.169  & 2.665$\pm$0.016 & 1.012$\pm$0.121 & 1.200$\pm$0.194 \\
M82 & 9.0 & 948.875$\pm$25.765  & 44.638$\pm$0.047 & 66.546$\pm$2.118 & 18.416$\pm$0.411 \\
NGC 3049 & 9.0 & 2.800$\pm$0.353  & 1.008$\pm$0.180 & 0.834$\pm$0.016 & 0.780$\pm$0.104 \\
NGC 1482 & 9.1 & 100.483$\pm$4.179  & 13.346$\pm$0.104 & 1.759$\pm$0.073 & 4.905$\pm$0.119 \\
NGC 2903 & 9.3 & 84.257$\pm$1.871  & 4.456$\pm$0.026 & 7.563$\pm$0.057 & 1.979$\pm$0.267 \\

  \enddata

\tablecomments{Columns: (1) Common source names; (2) Metallicity of the galaxy; (3) Extinction corrected flux of the 12.8 $\mu$m [NeII] fine
structure line in Watts per centimeter squared; (4) Extinction corrected flux of the 18.7 $\mu$m [SIII] fine structure line in Watts per centimeter
squared; (5) Extinction corrected flux of the 26.0 $\mu$m [FeII] fine structure line in Watts per centimeter squared; (6) Extinction corrected flux
of the 33.4 $\mu$m [SIII] fine structure line in Watts per centimeter squared.}
\end{deluxetable*}

\begin{deluxetable*}{lcccc}
\tabletypesize{\scriptsize} \tablecaption{IRS and optical flux ratios.} \tablewidth{0pt} \tablehead{\colhead{Target}    &
 \colhead{PAH 6.2 micron P/C}    &
 \colhead{[FeII]/[NeII]}     &
 \colhead{log ([SIII] 18.7/33.4 $\mu$m)} &
 \colhead{log ([SII] 6716/6731~\AA)}  \\
 \colhead{(1)}   &
 \colhead{(2)}   &
 \colhead{(3)}   &
 \colhead{(4)}   &
 \colhead{(5)}}

  \startdata

I Zw 18 & 1.2 & 1.74 & -0.29 & 1.32\tablenotemark{a} \\
SBS 0335-052 & 1.23 & 1.29 & -0.22 & 1.2\tablenotemark{b} \\
HS 2236+1344 & 1.88 & 1.75 & -0.25 & 1.15\tablenotemark{c} \\
SDSS J0519+0007 & 1.89 & 0.75 & 1.17\tablenotemark{c} \\
SBS 1415+437 & 1.57 & 1.11 & -0.24 & 1.15\tablenotemark{b} \\
HS 0837+4717 & 1.66 & 1.47 & -0.22 & 1.1\tablenotemark{c} \\
Tol 65 & 1.41 & 1.68 & -0.29 & 1.21\tablenotemark{c} \\
HS 1442+4250 & 1.81 & 0.56 & -0.20 & - \\
Mrk 25 & 2.56 & 0.60 & -0.11 & - \\
SBS 1030+583 & 1.67 & 0.37 & 0.05 & 1.25\tablenotemark{b} \\
Mrk 209 & 2.27 & 1.12 & -0.25 & 1.28\tablenotemark{a} \\
Mrk 36 & 1.82 & 0.46 & -0.23 & 1.07\tablenotemark{b} \\
SBS 0917+527 & 2.1 & 0.33 & 0.04 & 1.14\tablenotemark{b} \\
SBS 1152+579 & 2.41 & 0.33 & -0.14 & 1.17\tablenotemark{b} \\
Mrk 170 & 2.2 & 0.90 & -0.25 & \\
UM 448 & 1.63 & 0.31 & -0.15 & 1.07\tablenotemark{b} \\
SBS 0946+558 & 2.5 & 0.50 & -0.01 & 1.11\tablenotemark{b} \\
II Zw 40 & 2.6 & 0.26 & 0.03 & 1.15\tablenotemark{b} \\
Mrk 930 & 2.5 & 0.51 & -0.04 & 1.15\tablenotemark{b} \\
Mrk 996 & 3.3 & 0.24 & 0.17 & - \\
Tol 1924-416 & 2.8 & 0.13 & -0.04 & 1.13\tablenotemark{b} \\
II Zw 70 & 2.1 & 1.27 & -0.18 & 1.2\tablenotemark{b} \\
NGC 4670 & 3.75 & 0.64 & -0.13 & - \\
Mrk 450 & 2.64 & 0.40 & -0.07 & 1.17\tablenotemark{c} \\
Mrk 5 & 2.91 & 0.71 & -0.13 & 1.11\tablenotemark{b} \\
Mrk 1329 & 3.88 & 0.40 & 0.07 & 1.06\tablenotemark{c} \\
NGC 5253 & 3.21 & 0.71 & -0.13 & 0.91\tablenotemark{b} \\
IC 342 & 3.42 & 0.17 & 0.00 & 0.83\tablenotemark{e} \\
Haro 3 & 4.18 & 0.15 & 0.06 & 1.23\tablenotemark{a} \\
UM 311 & 3.27 & 0.44 & -0.06 & 0.92\tablenotemark{b} \\
Mrk 1499 & 4.27 & 0.46 & -0.09 & - \\
UGC 4274 & 3.92 & 0.43 & 0.01 & 0.9\tablenotemark{e} \\
NGC 7793 & 4.78 & 0.28 & 0.22 & - \\
NGC 4194 & 3.66 & 0.18 & 0.00 & - \\
NGC 253 & 3.05 & 0.14 & 0.17 & - \\
He 2-10 & 4.1 & 0.08 & 0.17 & 1.04\tablenotemark{f} \\
NGC 7714 & 4.22 & 0.12 & 0.35 & 0.88\tablenotemark{b} \\
M82 & 5.07 & 0.07 & 0.38 & 0.8\tablenotemark{e} \\
NGC 3049 & 5.23 & 0.30 & 0.11 & 0.87\tablenotemark{b} \\
NGC 1482 & 3.7 & 0.02 & 0.43 & - \\
NGC 2903 & 5.26 & 0.09 & 0.35 & 1.11\tablenotemark{e} \\

  \enddata

 \tablecomments{Columns:(1) Common source
names; (2) PAH 6.2 micron peak to continuum ratio; (3) [FeII]/[NeII] line ratio; (4) Logarithm of the [SIII] 18.7 to 33.4 micron line ratio; (5)
Logarithm of the [SII] 6716 to 6731~\AA~line ratio. Sources for [SII] ratios (a): \cite{kong99}; (b) \cite{hoyos06};  (c) \cite{izo04}; (d)
\cite{izo02}; (e) \cite{ho97}; (e) \cite{ho97}; (e) \cite{vacca92}.}
\end{deluxetable*}


\begin{thebibliography}{}

\bibitem[Abel et al.(2005)]{abel05}
Abel N.P., Ferland G.J., Shaw G. \& van Hoof P.A.M., 2005, \apjs, 161, 65

\bibitem[Anders \& Grevesse(1989)]{anders89}
Anders E. \& Grevesse N., 1989, Geochimica et Cosmochimica Acta, 53, 197

\bibitem[Bakes \& Tielens(1994)]{bakes94}
Bakes E.L.O. \& Tielens A.G.G.M., 1994, \apj, 427, 822

\bibitem[Baldwin et al.(1991)]{baldwin91}
Baldwin J.A., Ferland G.J., Martin P.G. et al., 1991 \apj, 374, 580

\bibitem[Brandl et al.(2004)]{bran04}
Brandl B.R., Devost D., Higdon S.J.U. et al., 2004, \apjs, 154, 188

\bibitem[Burkert(2004)]{burkert04}
Burkert A., 2004, The Formation and Evolution of Massive Young Star Clusters, ASP Conference Series, vol. 322., p.489

\bibitem[Calzetti et al.(2000)]{calzetti00}
Calzetti D., Armus L., Bohlin R.C. et al., 2000, \apj, 533, 682

\bibitem[Calzetti et al.(2004)]{calzetti04}
Calzetti D., Harris J., Gallagher J. et al., 2004, \aj, 127, 1405

\bibitem[Cardelli et al.(1989)]{cardelli89}
Cardelli J.A., Clayton G.C. \& Mathis J.S., 1989, \apj, 345, 245

\bibitem[Carral et al.(1994)]{carral94}
Carral P., Hollenbach D.J., Lord S.D. et al., 1994, \apj, 423, 223

\bibitem[Cazaux \& Tielens(2002)]{cazaux02}
Cazaux S. \& Tielens A.G.G.M., 2002, \apj, 575, 29


\bibitem[Dopita \& Sutherland(1996)]{dopita96}
Dopita M.A. \& Sutherland R.S., 1996, \apjs, 102, 61

\bibitem[Dopita(2000)]{dopita00}
Dopita M.A., 2000, Ap\&SS, 272, 79


\bibitem[Draine(1989)]{draine89}
Draine B. T., 1989, Interstellar Extinction in the Infrared, European Space Agency, p.93.

\bibitem[Draine(2003)]{draine03}
Draine B.T., 2003, \araa, 41, 241

\bibitem[Dudik(2007)]{dudik07}
Dudik R.P., Weingartner J.C., Satyapal S. et al., 2007, \apj, 684, 71.

\bibitem[Ferland et al.(1998)]{ferland98}
Ferland G.J., Korista K.T., Verner D.A. et al., 1998, \pasp, 110, 761

\bibitem[F\"{o}rster Schreiber et al.(2004)]{fors04}
F\"{o}rster Schreiber, N. M., Roussel, H., Sauvage, M., Charmandaris, V., 2004, \aap, 419, 501

\bibitem[Galliano(2008)]{galliano08}
Galliano F., Dwek E. \& Chanial P., 2008, \apj, 672, 214

\bibitem[Guhathakurta \& Draine(1989)]{guhathakurta89}
Guhathakurta P. \& Draine B.T, 1989, \apj, 345, 230

\bibitem[Higdon et al.(2004)]{hig04}
Higdon S.J.U., Devost D., Higdon J.L. et al., 2004, \pasp, 116, 975

\bibitem[Hogg et al.(2005)]{hogg05}
Hogg D.W., Tremonti C.A., Blanton M.R. et al., 2005, \aap, 388, 439

\bibitem[Hoyos \& Diaz(2006)]{hoyos06}
Hoyos C. \& Diaz A.I., 2006, \mnras, 365, 454

\bibitem[Ho et al.(1997)]{ho97}
Ho L.C., Filippenko A.V. \& Sargent W.L.W., 1997 \apjs, 112, 315

\bibitem[Izotov \& Thuan(2002)]{izo02}
Izotov Y.I. \& Thuan T.X., 2002, \apj, 567, 875

\bibitem[Izotov et al.(2004)]{izo04}
Izotov Y.I., Papaderos P., Guseva N.G. et al., 2004, \aap, 421, 539

\bibitem[Izotov et al.(2006)]{izotov06}
Izotov Y.I., Stasinska G., Meynet G. et al., 2006, \aap, 448, 955

\bibitem[Juett et al.(2006)]{juett06}
Juett A.M., Schulz N.S., Chakrabarty D., Gorczyca T.W., 2006, \apj, 648, 1066

\bibitem[Kendall \& Stuart(1976)]{ken76}
Kendall M. \& Stuart A. 1976, In: The Advanced Theory of Statistics, Vol. 2, Macmillian

\bibitem[Kewley et al.(2001)]{kewley01}
Kewley L.J., Dopita M.A., Sutherland R.S., Heisler C.A. \& Trevena J., 2001, \apj, 556, 121

\bibitem[Kewley \& Dopita(2002)]{kewley02}
Kewley L.J. \& Dopita M.A., 2002, \apjs, 142, 35

\bibitem[Kong \& Cheng(1999)]{kong99}
Kong X. \& Cheng F.Z., 1999, \aap, 351, 477

\bibitem[Leitherer et al.(1999)]{leitherer99}
Leitherer C., Schaerer D., Goldader J.D. et al., 1999, \apjs, 123, 3

\bibitem[Lord et al.(1996)]{lord96}
Lord S.D., Hollenbach D.J., Haas M.R. et al., 1996, \apj, 465, 703

\bibitem[Lutz, Veilleux \& Genzel(1999)]{lutz99}
Lutz D., Veilleux S., \& Genzel R., 1999, \apj 517, 13

\bibitem[Madden et al.(2006)]{mad06}
Madden S.C., Galliano F., Jones A.P. \& Sauvage, M., 2006, \aap, 446, 877

\bibitem[Mathis, Rumpl \& Nordsieck(1977)]{mat77}
Mathis J.S., Rumpl W. \& Nordsieck, K. H., 1977, \apj, 217, 425

\bibitem[O'Halloran, Satyapal \& Dudik(2006)]{oha06}
O'Halloran B., Satyapal S. \& Dudik R.P., 2006, \apj, 641, 795

\bibitem[Omont(1986)]{omont86}
Omont A., 1986, \aap, 164, 159

\bibitem[Osterbrock et al.(1992)]{osterbrock92}
Osterbrock D.E., Tran H.D. \& Veilleux S., 1992, \apj, 389, 305

\bibitem[Ott et al.(2005)]{ott05}
Ott J., Walter F. \& Brinks E., 2005, \mnras, 358, 1423

\bibitem[Reach \& Rho(2000)]{reach00}
Reach W.T. \& Rho J., 2000, \apj, 544, 843

\bibitem[Reach et al.(2006)]{reach06}
Reach W.T., Rho J., Tappe A. et al., 2006, \aj, 131, 479

\bibitem[Rigopoulou et al.(1996)]{rigo96}
Rigopoulou D., Lutz D., Genzel R. et al., 1996, \aap, 315, 125

\bibitem[Rigoupolou et al.(1999)]{rig99}
Rigopoulou, D., Spoon, H. W. W., Genzel, R. et al., 1999, \aj, 118, 2625

\bibitem[Rodriguez \& Rubin(2005)]{rodriguez05}
Rodriguez M. \& Rubin R.H., 2005, \apj, 626, 900

\bibitem[Rubin et al.(1991)]{rubin91}
Rubin R.H., Simpson J.P., Haas M.R. \& Erickson E.F., 1991, \apj, 374, 564

\bibitem[Satyapal et al.(2005)]{sat05}
Satyapal S., Dudik R.P., O'Halloran B. \& Gliozzi M., 2005, \apj, 633, 86

\bibitem[Scoville et al.(2000)]{sco00}
Scoville N.Z., Evans A.S., Thompson R. et al., 2000, \aj, 119, 991

\bibitem[Shaw et al.(2005)]{shaw05}
Shaw G., Ferland G.J., Abel N.P. et al., 2005, \apj, 624, 794

\bibitem[Skillman \& Bender(1995)]{skill95}
Skillman E.D. \& Bender R., 1995, RMxAC, 3, 255

\bibitem[Smith et al.(2005)]{smi05}
Smith B.J., Struck C. \& Nowak M.A., 2005, \aj, 129, 1350

\bibitem[Storey \& Hummer(1995)]{storey95}
Storey, P.J. \& Hummer D.G., 1995, \mnras, 272, 41

\bibitem[Strickland \& Stevens(2000)]{strickland00}
Strickland D.K. \& Stevens I.R., 2000, \mnras, 304, 511

\bibitem[Sturm et al.(2000)]{stu00}
Sturm E., Lutz D., Tran D. et al., 2000, \aap 358, 481

\bibitem[Surace \& Sanders(1999)]{sur99}
Surace J.A. \& Sanders D. B., 1999, \apj, 512, 162

\bibitem[Sutherland \& Dopita(1993)]{sutherland93}
Sutherland R.S. \& Dopita M.A., 1993, \apjs, 88, 253

\bibitem[Thomas et al.(2002)]{tho02}
Thomas H.C., Dunne L., Clemens, M.C. et al., 2002, \mnras, 329, 747

\bibitem[Vacca \& Conti(1992)]{vacca92}
Vacca W. \& Conti P., 1992, \apj, 401, 543

\bibitem[van Hoof et al.(2004)]{vanhoof04}
van Hoof P.A.M., Weingartner J.C., Martin P.G. et al., 2004, \mnras, 350, 1330

\bibitem[Verma et al.(2004)]{ver04}
Verma A., Lutz D., Sturm E. et al., 2004, \aap, 403, 829

\bibitem[Veilleux \& Osterbrock(1987)]{veilleux87}
Veilleux S. \& Osterbrock D.E., 1987, \apjs, 63, 295

\bibitem[Weingartner \& Draine(2001)]{weingartner01}
Weingartner J.C. \& Draine B.T., 2001, \apj, 548, 296

\bibitem[Wu et al.(2006)]{wu06}
Wu Y., Charmandaris V., Hao L. et al., 2006, \apj, 639, 157

\bibitem[Yao et al.(2006)]{yao06}
Yao Y., Schulz N., Wang Q.D. \& Nowak M., 2006, \apj, 653, 121
\end{thebibliography}
\end{document}